\documentstyle[amsfonts,epsfig,12pt]{article}

\def\mypagenumber{1}

\def\myend{\end{document}}

\normalsize

\newcounter{sxn}

\newcounter{axn}

\date{}

\newdimen\mybaselineskip
\newcommand{\beeq}{\begin{equation}}
\newcommand{\eneq}{\end{equation}}
\newcommand{\be}{\begin{eqnarray}}
\newcommand{\ee}{\end{eqnarray}}
\newcommand{\bpic}{\begin{picture}}
\newcommand{\epic}{\end{picture}}

\def\la{\raise.16ex\hbox{$\langle$} \, }
\def\ra{\, \raise.16ex\hbox{$\rangle$} }

\def\psibar{ \psi \kern-.65em\raise.6em\hbox{$-$} }
\def\mbar{ m \kern-.78em\raise.4em\hbox{$-$}\lower.4em\hbox{} }

\def\a{ {\alpha} }

\def\b{ {\beta} }

\def\n@space{\nulldelimiterspace=0pt \mathsurround=0pt }
\def\huge#1{{\hbox{$\left#1\vbox to 20.5pt{}\right.\n@space$}}}

\def\myskip{\noalign{\kern 8pt}}
\def\myeqspace{\noalign{\kern 10pt}}

\def\boxit#1{$\vcenter{\hrule\hbox{\vrule\kern3pt
    \vbox{\kern3pt\hbox{#1}\kern3pt}\kern3pt\vrule}\hrule}$}
\def\bigbox#1{$\vcenter{\hrule\hbox{\vrule\kern5pt
     \vbox{\kern5pt\hbox{#1}\kern5pt}\kern5pt\vrule}\hrule}$}

\def\ignore#1{{}}

\begin{document}
\bibliographystyle{unsrt}
\footskip 1.0cm

\thispagestyle{empty}
\setcounter{page}{\mypagenumber}


\begin{flushright}{
BRX-TH-509\\}

\end{flushright}

\vspace{2.5cm}
\begin{center}
{\LARGE \bf { Energy in Generic Higher Curvature Gravity
Theories }}\\
\vskip 1 cm
{\large{S. Deser and Bayram Tekin  }}\footnote{e-mail:~
deser, tekin@brandeis.edu}\\
\vspace{.5cm}
{\it Department of  Physics, Brandeis University, Waltham, MA 02454,
USA}\\
\begin{abstract}
We define and compute the energy of higher curvature gravity theories
in arbitrary dimensions. Generically, these theories admit
constant curvature vacua (even in the absence of an explicit
cosmological constant), and asymptotically constant curvature
solutions with non-trivial energy properties. For concreteness, we
study quadratic curvature models in detail. Among them, the one
whose action is the square of the traceless Ricci tensor always
has zero energy, unlike conformal (Weyl) gravity. We also study
the string-inspired Einstein-Gauss-Bonnet model and show that both
its flat and  Anti-de-Sitter vacua are stable.
\end{abstract}

\end{center}

\vspace*{1.5cm}

\newpage

\normalsize
\baselineskip=23pt plus 1pt minus 1pt
\parindent=27pt
\vskip 4 cm

\section{Introduction}

Recently, de Sitter (dS) and Anti-de Sitter (AdS) spaces have
received renewed interest both in string theory (AdS/CFT
correspondence) and in cosmology where a positive cosmological
constant may have been observed. This motivates a detailed study
of energy about these vacua, for systems that also involve higher
curvature terms, such as naturally arise in string theory and
other quantum theories of gravity. In this paper, we define and
compute the global charges (especially energy) of asymptotically
constant (including zero) curvature space-times for generic
gravitational models.

In a recent Letter {\cite{dt}}, which summarized some of the
present work, we defined the global charges primarily in four
dimensional quadratic theories. In this paper we extend that
discussion in several directions: We first present a reformulation
of the original definition \cite{abbott} of conserved charges in
cosmological Einstein theory; then we derive the generic form of
the energy for quadratic gravity theories in $D$ dimensions and
specifically study the ghost-free low energy string-inspired
model: Gauss-Bonnet (GB) plus Einstein terms.  We also briefly
indicate how higher curvature models can be similarly treated.

We will demonstrate that among purely quadratic theories, the one
whose Lagrangian is the square of the traceless Ricci tensor has
zero energy for all $D$ about its asymptotically flat or
asymptotically constant curvature vacua, unlike for example
conformal (Weyl) gravity in $D$=4.

A definition of gauge invariant conserved (global) charges in a
diffeomorphism-invariant theory rests on  the `Gauss law' and the
presence of asymptotic Killing symmetries. More explicitly, in
{\it{any}} diffeomorphism-invariant gravity theory, a vacuum
satisfying the classical equations of motion is chosen as the
background relative to which excitations and any background
gauge-invariant properties (like energy) are defined. Two 
important model-independent features immediately arise when
charges are so constructed: Firstly, the vacuum
itself always has zero charge; secondly, the charges are always
expressible as surface integrals. As we shall show below, a
generic, implicit, formulation (independent of the gravity model
considered) is rather simple and straightforward, although in the
applications of this generic picture for specific gravity models
complications arise due to, {\it e.g.}, choice of correct vacuum,
existence of global symmetries and converting `volume' to
`surface' integrals. Historically, the first application of this
procedure was in Einstein's gravity for flat backgrounds with its
Poincar\'{e} symmetries (``ADM  mass'' \cite{adm} ). The second step
was  to the  (A)dS vacua of cosmological Einstein theory (``AD mass"
\cite{abbott}).

The outline of the paper is as follows: In Sec. 2, we revisit the AD
\cite{abbott} Killing charge for the cosmological Einstein theory
and the energies of its Schwarzschild-(A)dS (collectively ``SdS" ) 
solutions. Sec.\ 3 is devoted to the derivation and computation of the
Killing charges in generic quadratic theories (with or without
Einstein terms) as well as their various limits, particularly in
Einstein-GB models. In Sec.\ 4, we discuss the purely quadratic
zero energy theory constructed from the traceless Ricci tensor.
Sec. 5 includes our conclusions as well as some open questions.
The Appendix collects some formulae useful for linearization
properties of quadratic curvature terms about (A)dS backgrounds.

\section{Reformulation of AD energy}

In this section, we reformulate the AD construction \cite{abbott}
and obtain new and perhaps more transparent surface integrals for
energy in cosmological Einstein theory. One of the reasons for
revisiting the AD formulation is, as it will become clear, that in
the higher curvature models we shall study in detail, the only
non-vanishing parts of energy, for asymptotically Schwarzschild
(Anti) de-Sitter spaces come precisely from the AD integrals, but
with essential contributions from the higher terms.

First, let us recapitulate \cite{dt} how conserved charges arise
in a generic gravity theory, coupled to a covariantly conserved,
bounded, matter source $\tau_{\mu \nu}$
 \be \Phi_{\mu \nu}(g, R,
\nabla R, R^2,...) = \kappa \tau_{\mu \nu}, \label{generic}
 \ee
where $\Phi_{\mu \nu}$ is the ``Einstein tensor" of a  local,
invariant, but otherwise arbitrary, gravity action and $\kappa$ is
the coupling constant. Now decompose the metric into the sum of a
background $\bar{g}_{\mu \nu}$ (which solves (\ref{generic}) for
$\tau_{\mu \nu} = 0$) plus a (not necessarily small) deviation
$h_{\mu \nu}$,  that vanishes sufficiently rapidly at infinity,
\be
g_{\mu \nu}= \bar{g}_{\mu \nu}+ h_{\mu \nu}.
 \ee
Separating the field equations (\ref{generic}) into a part linear
in $h_{\mu \nu}$ plus all the non-linear ones that constitute the
total source $T_{\mu \nu}$, including the matter source $\tau_{\mu
\nu}$, one obtains
 \be  {\cal{O}}(\bar{g})_{\mu \nu
\alpha \beta}h^{\alpha \beta} = \kappa T_{\mu\nu}, \label{ope}
 \ee
as $\Phi_{\mu \nu}(\bar{g}, \bar{R},
\bar{\nabla}\bar{R},{\bar{R}}^2 ...)= 0$, by assumption;  the
hermitian operator ${\cal{O}}(\bar{g})$ depends only on the
background metric (that also moves all indices and defines the
covariant derivatives $\bar{\nabla}_{\mu}$). It is clear that this
operator inherits both background Bianchi identity and background
gauge invariance namely, $\bar{\nabla}_{\mu}{\cal{O}}(\bar{g})^{\mu
\nu \alpha \beta}= {\cal{O}}(\bar{g})^{\mu \nu \alpha
\beta}\bar{\nabla}_\a =0$, from (the Bianchi identities of) the
full theory. As a consequence of these invariances, it is
guaranteed that if the background $\bar{g}_{\mu\nu}$ admits a set
of Killing vectors $\bar{\xi}_{\mu}^{(a)}$,
 \be {\bar{\nabla}}_\mu
\bar{\xi}_\nu^{(a)} + {\bar{\nabla}}_\nu \bar{\xi}_\mu^{(a)}  = 0,
 \ee
 then the
energy-momentum tensor can be used to construct the following
(ordinarily) conserved vector density current,
 \be
\bar{\nabla}_{\mu} ( \sqrt{-\bar{g}} T^{\mu \nu}\bar{\xi}_\nu^a)
\equiv
\partial_\mu ( \sqrt{-\bar{g}} T^{\mu \nu}\bar{\xi}_\nu^a) = 0.
\ee 
Therefore, the conserved Killing charges are expressed as
 \be
Q^\mu(\bar{\xi}^a) = \int_{\cal{M}} d^{D-1} x \sqrt{-\bar{g}}
T^{\mu \nu}\bar{\xi}_\nu^a =  \int_{\Sigma} dS_i {\cal{F}}^{\mu i}
\; . \label{charge}
 \ee
Here ${\cal M}$ is a spatial ($D$--1) hypersurface and 
$\Sigma$ is its 
($D-2$) dimensional boundary; ${\cal{F}}^{\mu i}$ is an antisymmetric
tensor obtained from ${\cal{O}}(\bar{g})$, whose explicit form, of
course, depends on the theory.

Let us first apply the above procedure to
cosmological Einstein theory to rejoin \cite{abbott}. Our
conventions are:  signature $(-,+,+, ... +)$,
$[\nabla_\mu, \nabla_\nu]V_{\lambda} =
R_{\mu \nu \lambda}\,^\sigma V_\sigma $,\,\,
$ R_{\mu \nu} \equiv R_{\mu \lambda \nu}\,^\lambda$. The 
Einstein equations 
\be
R_{\mu \nu} - {1 \over 2} g_{\mu \nu} R + \Lambda g_{\mu \nu} = 0,
\label{einstein}
\ee
are solved by the 
constant curvature vacuum $\bar{g}_{\mu \nu}$, whose
Riemann, Ricci and scalar curvature are 
\be
\bar{R}_{\mu \lambda \nu \beta } = { 2\over(D-2)(D-1)}
\Lambda ( \bar{g}_{\mu \nu} \bar{g}_{\lambda \beta} -
\bar{g}_{\mu \beta} \bar{g}_{\lambda \nu } )
\hskip 0.5 cm
\bar{R}_{\mu \nu}= {2 \over {D- 2}}\Lambda \bar{g}_{\mu \nu}, \hskip 0.5 cm
\bar{R} = {2 D\Lambda \over {D-2}}.
\label{symmetric}
\ee
Linearization of (\ref{einstein})
about this background yields
\be
{\cal{G}}^L_{\mu \nu} \equiv
R_{\mu \nu}^L - {1\over 2} \bar{g}_{\mu \nu} R^L -
{2 \over D-2}\Lambda h_{\mu \nu} \equiv \kappa T_{\mu \nu},
\label{lineins}
\ee
where $R^L= (g^{\mu \nu}R_{\mu \nu})_L$ and
the linear part of the Ricci tensor reads
\be
R_{\mu \nu}^L \equiv  R_{\mu \nu} - \bar{R}_{\mu \nu} =
{1\over 2}( - {\bar{\Box}} h_{\mu\nu} -
{\bar{\nabla}}_\mu {\bar{\nabla}}_\nu h  +
{\bar{\nabla}}^{\sigma}{\bar{\nabla}}_\nu h_{\sigma \mu} +
{\bar{\nabla}}^{\sigma}{\bar{\nabla}}_\mu h_{\sigma \nu}),
\label{linearricci}
\ee
with $h = \bar{g}^{\mu \nu}h_{\mu \nu}$ and
$\bar{\Box} = \bar{g}^{\mu \nu}\bar{\nabla}_{\mu} \bar{\nabla}_\nu$.
The energy momentum-tensor (\ref{lineins})
is background covariantly constant ($\bar{\nabla}_\mu T^{\mu \nu} =0$), as
can be checked explicitly.

This procedure led \cite{abbott} to the following
energy expression 
\be
E(\bar{\xi}) = {1\over 8\pi G}\int dS_i \sqrt{-\bar{g}}
\left \{ \bar{\xi}_\nu \bar{\nabla}_{\beta} K^{0i\nu\beta} - K^{0j\nu i}
\bar{\nabla}_j \bar{\xi}_\nu \right \}.
\ee
The superpotential $K^{\mu \alpha \nu \beta}$ is defined by
\be
K^{\mu \nu \alpha \beta} \equiv {1\over 2} [ \bar{g}^{\mu \beta}H^{\nu \alpha}
+ \bar{g}^{\nu \alpha}H^{\mu \beta} - \bar{g}^{\mu \nu}H^{\alpha \beta} -
\bar{g}^{\alpha \beta}H^{\mu \nu}], \hskip 1 cm H^{\mu \nu} =
h^{\mu \nu} - {1\over 2}\bar{g}^{\mu \nu}h.
\ee
It has the symmetries of the Riemann tensor.
In converting the volume to surface integrals, we adopt here,
a somewhat, different route, which will be convenient in
the higher curvature cases. Using 
(\ref{lineins}), (\ref{linearricci}), straightforward rearrangements 
of terms, and  the aforementioned 
antisymmetry, we  can move the covariant
derivatives to yield
\be
&&2 \bar{\xi}_\nu {\cal{G}}^{\mu \nu}_L =
2\bar{\xi}_\nu R^{\mu \nu}_L - \bar{\xi}_\nu \bar{g}^{\mu \nu} R^L -
{4\Lambda \over D-2}\bar{\xi}_\nu h^{\mu \nu}\nonumber \\
&& = \bar{\xi}_\nu \Big \{- {\bar{\Box}} h^{\mu\nu} -
{\bar{\nabla}}^\mu {\bar{\nabla}}^\nu h  +
{\bar{\nabla}}_{\sigma}{\bar{\nabla}}^\nu h^{\sigma \mu} +
{\bar{\nabla}}_{\sigma}{\bar{\nabla}}^\mu h^{\sigma \nu}\Big \}
-\bar{\xi}^\mu \Big \{- \bar{\Box} h +
\bar{\nabla}_\sigma \bar{\nabla}_\nu
h^{\sigma \nu} - {2\Lambda\over D-2} h \Big \} \nonumber \\
&& \hskip 1 cm - {4\Lambda \over D-2}\bar{\xi}_\nu h^{\mu \nu} \hskip 0.3 cm
= \hskip 0.3 cm \bar{\nabla}_\rho \Big \{
\bar{\xi}_\nu \bar{\nabla}^{\mu}h^{\rho \nu} -\bar{\xi}_\nu
\bar{\nabla}^{\rho}h^{\mu\nu}
+\bar{\xi}^\mu \bar{\nabla}^\rho h
-\bar{\xi}^\rho \bar{\nabla}^\mu h \nonumber \\
&& \hskip 2 cm + h^{\mu \nu}\bar{\nabla}^\rho \bar{\xi}_\nu
- h^{\rho \nu}\bar{\nabla}^\mu \bar{\xi}_\nu +
\bar{\xi}^\rho \bar{\nabla}_{\nu}h^{\mu \nu}
-\bar{\xi}^\mu \bar{\nabla}_{\nu}h^{\rho \nu} +
h\bar{\nabla}^\mu \bar{\xi}^\rho \Big \}.
\ee
Having now rewritten all as a surface term,
the Killing charges of  \cite{abbott} become
\be
Q^{\mu}(\bar{\xi}) = {1\over 4 \Omega_{D-2} G_D}\int_{\Sigma} &dS_i&\Big \{
\bar{\xi}_\nu \bar{\nabla}^{\mu}h^{i \nu} -\bar{\xi}_\nu
\bar{\nabla}^{i}h^{\mu\nu}
+\bar{\xi}^\mu \bar{\nabla}^i h
-\bar{\xi}^i \bar{\nabla}^\mu h \nonumber \\
&&+h^{\mu \nu}\bar{\nabla}^i \bar{\xi}_\nu
- h^{i \nu}\bar{\nabla}^\mu \bar{\xi}_\nu +
\bar{\xi}^i \bar{\nabla}_{\nu}h^{\mu \nu}
-\bar{\xi}^\mu \bar{\nabla}_{\nu}h^{i \nu} + h\bar{\nabla}^\mu
\bar{\xi}^i \Big \},
\label{ad}
\ee
where $i$ ranges over ($1, 2,... D-2$); the charge is normalized by
dividing by the (D-dimensional ) Newton's constant and solid angle.
Before we perform the explicit computation of the energy $Q^0$ in
specific coordinates for asymptotically (A)dS spaces,
let us check  that this charge is in fact background gauge-invariant.
Under an infinitesimal diffeormophism, generated by a vector
$\zeta_{\mu}$, the deviation part of the metric
transforms as
\be
\delta_\zeta h_{\mu \nu} = \bar{\nabla}_\mu \zeta_\nu +
\bar{\nabla}_\nu \zeta_\mu.
\ee
To show that $T_{\mu \nu}\bar{\xi}^\nu$ is invariant, first 
note that $R_L$ is:
\be
\delta_\zeta R_L = \bar{g}^{\mu \nu} \delta_\zeta R_{\mu \nu}^L -
{2\over D-2} \Lambda  \bar{g}^{\mu \nu}\delta_\zeta h_{\mu \nu} = 0.
\ee
This leads to $\delta_\zeta {\cal{G}}_{\mu \nu}^L = {2\over D-2}\Lambda
\delta_\zeta h_{\mu \nu}$ and eventually to $\delta_\zeta Q^\mu = 0$:
the Killing charge is indeed background gauge-invariant.
Another test of (\ref{ad}) is that,
in the limit of an asymptotically flat background, we should obtain the
ADM charge. In that case, we may write  the time-like Killing 
vector as $\xi_{\mu} =
(1,{\bf{0}})$.
The time component of (\ref{ad}), reduces to the desired
result,
\be
Q^0 = M_{ADM} = {1\over 4 \Omega_{D-2} G_D}\int_{\Sigma} &dS_i&\Big \{
\partial_j h^{ij} - \partial^i h_{jj} \Big \}
\ee
in terms of Cartesian coordinates.

Having established the energy formula for asymptotically (A)dS
spaces, we can now evaluate the energy of SdS solutions.
First, we must recall that  the existence of a cosmological 
horizon is an important difference
between dS and AdS cases. In the former, the background Killing
vector stays time-like only within the cosmological horizon.
[We will not go into the complications for physics of this horizon,
since it is a well-known and ongoing problem. In \cite{abbott}, 
it was simply assumed that interesting system should be
describable within the horizon. For related ideas see \cite{witten}.]
For small black holes,
whose own event horizons lie well inside the cosmological one, (\ref{ad})
provides a reasonable approximation.

In static coordinates, the line element of  $D$-dimensional SdS reads
\be
ds^2 = - \Big \{ 1- ({ r_0\over r})^{D-3} - {r^2\over l^2}\Big \}\, dt^2 +
 \Big \{1- ({ r_0\over r})^{D-3} -  {r^2\over l^2}\Big \}^{-1}\,dr^2 +
r^2 d\Omega_{D-2}^2,
\label{sds}
\ee
where $l^2 \equiv {(D-2)(D-1)\over 2\Lambda}$. The background 
($r_0 =0$) Killing vector is  $\xi^\mu = ( -1, {\bf{0}} )$,
which is time-like everywhere for AdS ($l^2  < 0$)
but remains time-like for dS ($l^2 >0 $) only inside the cosmological
horizon: $\bar{g}_{\mu \nu}{\bar{\xi}}^\mu \bar{\xi}^\nu = -
( 1 - {r^2\over l^2})$.

Let us concentrate on $D=4$ first and calculate
the surface integral ({\ref{ad}) not at $r = \infty$, but at 
some finite
distance $r$ from the origin; this  will {\it{not}} be
gauge-invariant, since energy is to be measured  
only at infinity.
Nevertheless, for dS space (which
has a horizon that keeps us from going smoothly to infinity), 
let us first keep $r$ finite as an
intermediate step. The integral becomes
\be
E(r) = {r_0\over 2 G}{ (1- {r^2\over l^2}) \over (1- {r_0\over r} -
{r^2\over l^2})}.
\label{finiter}
\ee
For AdS, 
$E( r \rightarrow \infty) = {r_0 \over 2 G } \equiv M$, as expected.
On the other hand, for dS $E(r = l) =0$. 
This is, however, misleading, since, in dS we should really only 
consider small $r_0$ objects, which do not 
change the location of the background
horizon. [Indeed, if we naively include the effect of a
large $r_0$ as changing the horizon to  $ 1- {r_0\over r} -
{r^2\over l^2}= 0$, then $E(r)$ itself diverges!.] 
But, as stated, we derived the energy formula  using 
asymptotic Killing vectors. Therefore, for asymptotically dS spaces, 
the only way to make
sense of the above result is to consider the small $r_0$ limit, which then
gives $ E= M$ \cite{abbott}. In the limit of
a vanishing cosmological constant (namely $l \rightarrow \infty$ ), 
the ADM energy is of course recovered as  $r \rightarrow \infty$.

The above argument easily generalizes to $D$-dimensions, where one
obtains
\be
E= {D-2 \over 4 G_D}r_0^{D-3}.
\ee
Here $r_0$ can be arbitrarily large in the AdS case
but must be small in dS.

Finally, let us note that analogous computations can also be 
carried out in $D=3$; the proper solution is
\be
ds^2 = -( 1- r_0 - {r^2\over l^2} )dt^2 +
( 1- r_0 - {r^2\over l^2})^{-1}dr^2 + r^2 d\phi^2
\label{3dmetric}
\ee
for which  the energy is $E = r_0/ 2G$ again but, now, $r_0$ is a dimensionless
constant and $[G] = [M^{-1}]$, in agreement with the 
original results \cite{jackiw}.

\section{ String-inspired Gravity }

In flat backgrounds, the
ghost-freedom of low energy string theory  requires the
quadratic corrections to Einstein's gravity to be of
the Gauss-Bonnet (GB) form \cite{zwiebach}, an argument that
should carry over to the AdS backgrounds.
Below we construct and compute
the energy of various asymptotically (A)dS spaces that solve
generic Einstein plus quadratic gravity theories, particularly,
Einstein-GB model.

At quadratic order, the generic action is~\footnote{We will later add
an explicit cosmological constant term in the discussion. Note
also that the normalizations of $\a, \b$ differ from those
of \cite{dt}.}
\be
I = \int d^D\, x \sqrt{-g} \Big \{ {R\over \kappa} +
\a R^2 + \b R^2_{\mu\nu} +
\gamma (R_{\mu\nu\rho\sigma}^2 -4R^2_{\mu\nu}+R^2 ) \Big \}.
\label{action}
\ee
In $D = 4$, the GB part ($\gamma$ terms ) is a
surface integral and plays no role in the equations of motion.
In $D >4$,
on the contrary, GB is the only  viable term,
since non-zero $\a$, $\b$ produce ghosts \cite{stelle}. Here 
$\kappa =
2 \Omega_{D-2} G_D $, where $G_D$ is the D-dimensional Newton's constant.

The equations of motion that follow from (\ref{action}) are
\be
&& {1\over \kappa}(R_{\mu \nu} -{1\over 2}g_{\mu \nu} R)
+ 2\alpha R\,(R_{\mu\nu} - {1\over 4}g_{\mu\nu}\,R ) +
(2\alpha +\beta) (g_{\mu\nu}\Box - \nabla_\mu \nabla_\nu )R \nonumber \\
&&+ 2\gamma \Big \{ R R_{\mu \nu} - 2 R_{\mu\sigma\nu\rho}R^{\sigma \rho}
+ R_{\mu\sigma\rho\tau}R_\nu^{\sigma\rho\tau} - 2R_{\mu \sigma}R_\nu^\sigma
- {1\over 4}g_{\mu \nu}(R_{\tau\lambda\rho\sigma}^2 -4R^2_{\sigma\rho}+R^2 )
\Big \} \nonumber \\
&& +\beta \Box ( R_{\mu\nu} - {1\over 2}g_{\mu\nu} R)
+2\beta ( R_{\mu \sigma \nu \rho} -
{1\over 4}g_{\mu \nu}R_{\sigma \rho})R^{\sigma \rho} = \tau_{\mu \nu}.
\label{eom}
\ee
In the absence of matter, flat space is {\it{a}} solution of these
equations. But so is (A)dS with cosmological constant 
$\Lambda$, which in our conventions is (see also {\cite{odintsov}})
\be
-{1\over 2 \Lambda \kappa} = { (D-4)\over (D-2)^2}(D\a +\b) +
{\gamma (D-4)(D-3)\over (D-2)(D-1) } \hskip 1 cm  \Lambda \ne 0.
\label{lam1}
\ee
Several comments are in order here. In the string-inspired EGB model,
($\alpha =\beta =0$ and $\gamma  > 0$), 
{\it{only}} AdS background ($\Lambda < 0$ ) is allowed
(the Einstein constant $\kappa$ is positive with our conventions).
String theory is known to prefer AdS to dS 
( see for example the no-go theorem \cite{nunez} )
we can now see why this is so in the uncompactified
theory. Another interesting limit is the `traceless' theory
( $D\a = -\b$), which, in the absence of a $\gamma$ term, does not
allow constant curvature spaces unless the Einstein term is also dropped.
For $D = 4$, the $\gamma$ term drops out, and the pure quadratic
theory allows (A)dS solutions with arbitrary $\Lambda$.
For $D >4$, relation (\ref{lam1}) leaves  a 2-parameter
set (say $\a, \b$ ) of allowed solutions for chosen $(\kappa, \Lambda)$ 
just like in $D=4$.

Following the procedure outlined in the previous section and using
the formulae in the Appendix, we expand the
field equations to first order in $h_{\mu \nu}$ and define the total
energy momentum tensor as
\be
T_{\mu \nu}(h) &\equiv& T_{\mu \nu}(\bar{g})+ {\cal{G}}^L_{\mu\nu} \Big \{
{1\over \kappa } + { 4\Lambda D\a \over D-2} + {4\Lambda\b\over D-1} +
{4\Lambda \gamma (D-4)(D-3)\over (D-2)(D-1)} \Big \}\nonumber \\
&&+(2\alpha +\beta)({\bar{g}}_{\mu\nu}\bar{\Box} -
{\bar{\nabla}}_\mu {\bar{\nabla}}_\nu +{2\Lambda\over D-2} g_{\mu \nu})R_L
+\beta ( \bar{\Box} {\cal{G}}^L_{\mu\nu} -
{2\Lambda \over D-1} \bar{g}_{\mu \nu}R_L)\nonumber \\
&& - 2\Lambda^2 h_{\mu \nu}\Big \{
{1\over 2 \Lambda \kappa} + {(D-4)\over (D-2)^2}(D\a +\b) +
{\gamma (D-4)(D-3)\over (D-2)(D-1)} \Big \}.
\label{totalT}
\ee
Using (\ref{lam1}) one has $ T_{\mu \nu}(\bar{g}) = 0$ and the last term
also vanishes, yielding
\be
T_{\mu \nu} &=& {\cal{G}}^L_{\mu\nu} \Big \{
- {1\over \kappa} + { 4\Lambda D\over (D-2)^2}( 2\a  +{\b\over D-1})
\Big \}\nonumber \\
&&+(2\alpha +\beta)({\bar{g}}_{\mu\nu}\bar{\Box} -
{\bar{\nabla}}_\mu {\bar{\nabla}}_\nu +{2\Lambda\over D-2} g_{\mu \nu})R_L
+\beta ( \bar{\Box} {\cal{G}}^L_{\mu\nu} -
{2\Lambda \over D-1} \bar{g}_{\mu \nu}R_L).
\label{finalT}
\ee
This is a background conserved tensor (
$\bar{\nabla}^\mu T_{\mu \nu} = 0$ ) as can be checked explicitly with help
of the expressions
\be
&&\bar{\nabla}^{\mu}({\bar{g}}_{\mu\nu}\bar{\Box} -
{\bar{\nabla}}_\mu {\bar{\nabla}}_\nu
+{2\Lambda\over D-2} g_{\mu \nu})R_L =0 \nonumber \\
&&\bar{\nabla}^{\mu}( \bar{\Box} {\cal{G}}^L_{\mu\nu} -
{2\Lambda \over D-1} \bar{g}_{\mu \nu})R_L=0.
\ee
An important aspect of (\ref{finalT}) is the sign change
of the $1/\kappa$ term relative to Einstein theory, due
to the GB contributions 
as already noticed in \cite{boulware}. Hence in the Einstein-GB
limit, we have
$T_{\mu \nu} = - {\cal{G}}^L_{\mu\nu}/\kappa$, with 
overall sign exactly opposite~\footnote{This overall sign change
is also shared by the model's small oscillations about the AdS vacuum.} 
to that of the cosmological Einstein theory (\ref{lineins}). But, 
as we shall see below,
this does not mean that $E$ is  negative there.

There remains now to obtain a Killing energy expression from
(\ref{finalT}), namely, to write
$\bar{\xi}_\nu T^{\mu \nu}$ as a surface integral.
The first term is the usual AD piece ({\ref{ad}),
which we have already dealt with
in the previous section. The middle term with the coefficient
$ 2\a + \b$, is easy to handle. The relatively cumbersome last term can be
written as a surface plus extra terms.
\be
\bar{\xi}_\nu \bar{\Box} {\cal{G}}^L_{\mu\nu} &=&
\bar{\nabla}_\alpha \Big \{\bar{\xi}_\nu
\bar{\nabla}^{\alpha} {\cal{G}}_L^{\mu \nu}
- \bar{\xi}_\nu \bar{\nabla}^{\mu} {\cal{G}}_L^{\alpha \nu} -
{\cal{G}}_L^{\mu\nu} \bar{\nabla}^{\alpha}
\bar{\xi}_\nu + {\cal{G}}_L^{\alpha\nu}\bar{\nabla}^{\mu}\bar{\xi}_\nu
\Big \} \nonumber \\
&& + {\cal{G}}_L^{\mu \nu}\Box \bar{\xi}_\nu +
\bar{\xi}_\nu \bar{\nabla}_\alpha \bar{\nabla}^\mu {\cal{G}}^{\alpha \nu}
-{\cal{G}}_L^{\alpha \nu} \bar{\nabla}_\alpha \bar{\nabla}^\mu \bar{\xi}_\nu
\ee
Using the definition of the Killing vector, and its trace property,
\be
\bar{\nabla}_\alpha \bar{\nabla}_\beta \bar{\xi}_\nu =
\bar{R}_{\nu\beta\alpha}^\mu \bar{\xi}_{\mu} =
{2\Lambda \over (D-2)(D-1)}( \bar{g}_{\nu \alpha}\bar{\xi}_\beta
-\bar{g}_{\alpha \beta }\bar{\xi}_\nu  ), \hskip 0.5 cm
\Box \bar{\xi}_\mu = -{2\Lambda \over D-2}\bar{\xi}_\mu,
\ee
along with the identity
\be
\bar{\xi}_\nu\bar{\nabla}_\alpha \bar{\nabla}^\mu {\cal{G}}^{\alpha \nu}_L =
{2\Lambda D\over (D-2)(D-1)}\bar{\xi}_\nu {\cal{G}}^{\mu \nu}_L +
{\Lambda \over D-1}\xi^\mu R_L,
\ee
one can show that $\bar{\xi}_\nu \bar{\Box} {\cal{G}}^L_{\mu\nu}$
can indeed be written as a surface term. 
Collecting everything, the final form
of the conserved charges for the generic quadratic theory reads
\be
Q^\mu(\bar{\xi}) &=&
\Big \{-{1\over \kappa}+ {8\Lambda\over (D-2)^2}(D\alpha +\beta) \Big \}
\int d^{D-1}x\, \sqrt{ -\bar{g}}
\bar{\xi_\nu}{\cal{G}}_L^{\mu\nu}  \nonumber \\
&&+ (2\alpha +\beta)
\int  dS_i \sqrt{-g} \Big \{ \bar{\xi}^\mu
\bar{\nabla}^i R_L + R_L \bar{\nabla}^\mu\, \bar{\xi}^i
- \bar{\xi}^i \bar{\nabla}^\mu R_L \Big \} \nonumber \\
&& +\beta \int dS_i \sqrt{-g}\Big \{ \bar{\xi}_\nu
\bar{\nabla}^{i} {\cal{G}}_L^{\mu \nu}
- \bar{\xi}_\nu \bar{\nabla}^{\mu} {\cal{G}}_L^{i \nu} -
{\cal{G}}_L^{\mu\nu}
\bar{\nabla}^{i} \bar{\xi}_\nu + {\cal{G}}_L^{i \nu}
\bar{\nabla}^{\mu}\bar{\xi}_\nu \Big \}.
\label{fullcharge}
\ee
For brevity we have left the AD part as a volume integral whose surface form
we know is given by (\ref{ad}); note that $\gamma$ does not appear explicitly
since it has been traded for $\Lambda$ through the relation (\ref{lam1}).

In the above analysis, there was no bare
cosmological term in the action. Clearly, this need not be the case:
we can add one, say  : $2\int d^D x \sqrt{- g} \Lambda_0/ \kappa$.
The $\Lambda_0$
contributes to the overall
{\it{effective}} cosmological constant $\Lambda$, which
now is given by 
\be
\Lambda = -{1\over 4 f(\alpha,\beta, \gamma_) \kappa} \left \{ 1 \pm
\sqrt{ 1 + 8 \kappa f(\alpha,\beta, \gamma) \Lambda_0} \right \}
\ee
\be
f(\alpha, \beta, \gamma ) \equiv { (D-4)\over (D-2)^2}(D\a +\b) +
{\gamma (D-4)(D-3)\over (D-2)(D-1) }. \nonumber
\ee
If  $f > 0$, as in Einstein-GB theory, the effective cosmological constant
$\Lambda$ is smaller than the `bare' one $\Lambda_0$: thus stringy
corrections (at quadratic order) reduce the value of the
bare cosmological constant appearing in the Lagrangian. Given that
$\Lambda_0$ is arbitrary, there is a bound ($ 8 \kappa\Lambda_0 f \ge -1 $)
on these corrections since the effective $\Lambda$ becomes imaginary
otherwise.

Now let us compute the energy of an asymptotically SdS geometry
that might be a solution to our generic model. 
Should such a solution exist, we only require its asymptotic 
behavior to be 
\be
h_{0 0} \approx + ({ r_0\over r})^{D-3}, \hskip 1 cm h^{rr}  \approx
+({ r_0\over r})^{D-3} + O(r_0^2).
\ee
It is easy to see that for asymptotically SdS spaces
the second and the third lines of (\ref{fullcharge})
do not contribute, since for any Einstein space, to linear order
\be
R_{\mu \nu}^L ={2 \Lambda \over D-2} h_{\mu \nu},
\ee
which in turn yields $R_L = \bar{g}^{\mu \nu}R_{\mu \nu}^L -{2\Lambda\over
D-2} h= 0 $ and thus ${\cal{G}}_{\mu \nu}^L = 0 $
in the asymptotic region. Therefore the total energy of the full
$(\a, \b, \gamma)$ system, for geometries which are asymptotically SdS, is
given only by the first term in (\ref{fullcharge}),
\be
E_D = \left \{ -1 + {8 \Lambda \kappa\over (D-2)^2}(D\alpha +\beta)
\right \} {(D-2)\over 4 G}r_0^{D-3}
\label{adsen} \hskip 0.5 cm D > 4,
\ee
where $\gamma$ is implicitly assumed not to vanish. [Note 
again the sign change of the ``Einstein contribution as explained
before.] For  $D = 4$, we computed $E$ in {\cite{dt}},
equivalently from (\ref{totalT}), it reads
~\footnote{ In $D=3$, the GB density vanishes  identically 
and the energy expression has the same form of the $D=4$ model, with
the difference that $r_0$ comes from the metric
(\ref{3dmetric})}
 ( for models with an explicit
$\Lambda$)
\be
E_4 =\left \{ 1 + 2 \Lambda \kappa (4\alpha +\beta)\right \}
{r_0\over  2 G}.
\label{4ad}
\ee
From (\ref{adsen}), the asymptotically
SdS solution seemingly has negative energy, in the Einstein-GB model:
\be
E = -{(D-2)\over 4 G}r_0^{D-3}.
\ee
While this is of course correct in terms of the 
usual SdS signs, one must be more careful about the external
solutions in Einstein-GB theory. Their exact form is
{\cite{boulware}},
\be
ds^2 = g_{00}dt^2 + g_{rr}dr^2 + r^2 d\Omega_{D-2}
\ee
\be
-g_{00} = g^{-1}_{rr} = 1+ {r^2\over 4\kappa \gamma(D-3)(D-4)}
\left \{ 1 \pm
\left \{ 1 + 8\gamma 
(D-3)(D-4){r_0^{D-3}\over r^{D-1}}\right \}^{1\over2}\right \}
\label{exactsolution}
\ee
Note that there is a branching here, with qualitatively different
asymptotics: Schwarzschild and
Schwarschild-AdS,
\be
-g_{00} = 1- ({r_0\over r})^{D-3}, \hskip 1 cm -g_{00} =
1 +({r_0\over r})^{D-3} +{r^2\over \kappa \gamma(D-3)(D-4)}.
\label{ast}
\ee
[ Here  we have restored $\gamma$, using  
$\kappa \gamma(D-3)(D-4)= - l^2$.]
The first solution has the usual positive (for
positive $r_0$ of course) ADM  energy
$E = +{(D-2)r_0^{D-3}\over 4 G}$, since the GB term does not
contribute when expanded around flat space.
On the other hand,
as noted in {\cite{boulware}} the second solution which is asymptotically
SdS, has the wrong sign for the `mass term'. But, to actually
compute the energy here, one needs our energy
expression (\ref{fullcharge}), and
not simply the AD formula which is valid only for cosmological Einstein
theory. Now from (\ref{ast}), we have
\be
h_{00} \approx - ({ r_0\over r})^{D-3}, \hskip 1 cm h^{rr}
\approx  -({ r_0\over r})^{D-3} + O(r_0^2),
\ee
whose sign is opposite to that of
the usual SdS. This sign just compensates the 
flipped sign in the energy definition, so the energy (\ref{adsen})
reads : $E = {(D-2)r_0^{D-3}\over 4 G}$ and the
AdS branch, just like the flat branch, has positive energy, after
the GB effects are taken into account also in the energy definition.
Thus, for every Einstein-GB external solution, energy is positive and
AdS vacuum is stable.~\footnote{ In \cite{boulware}, 
it was erroneously concluded that
$E_D$ was negative for the AdS branch, despite having
obtained both the correct (negative) sign of $T^{\mu \nu}$ and 
of course the correct solution (\ref{exactsolution}).}

\section{Zero Energy Models}

In $D=4$, every quadratic curvature theory, {\it{i.e}} any $(\a,\b)$
combination, is scale invariant. These models were studied 
in \cite{strominger} in terms of 
 the slightly different
parametrization
\be
S = \int d^4 x \sqrt{-g} \left \{ a C_{\mu \nu \sigma \rho}\,
C^{\mu \nu \sigma \rho} + b R^2 \right \}
\ee
where $C_{\mu \nu \sigma \rho}$ is the Weyl tensor. Using the equivalent
of the ADM energy for the asymptotically flat solutions, it was shown
that this energy vanished for all of them.
As discussed in \cite{dt}, with our definition of energy, this
statement is correct, but simply reflects (at Einsteinian level) 
the Newtonian impossibility of having  asymptotically vanishing solutions 
of $\nabla^4 \phi = \rho$. This property of higher derivative 
gravity is well-understood \cite{havas}. It has 
deeper consequences such as violations of the equivalence principle: 
massive sources here have no gravitational mass. Violations 
of the equivalence principle are not unheard-of and occur already
at the simple level of scalar-tensor gravity. In the asymptotically 
(A)dS branch, however, energy no longer vanishes: Even 
pure conformally invariant Weyl theory has finite energy!

Interestingly, there is one purely quadratic theory which {\it{does}}
have vanishing energy in {\it{all}} dimensions,
for asymptotically flat or (A)dS vacua. It has action 
$\int d^D x\sqrt{-g} \tilde{R}_{\mu \nu} \tilde{R}^{\mu \nu}$,
where $\tilde{R}_{\mu \nu} \equiv  R_{\mu \nu} - {R\over D} g_{\mu \nu}$.
This vanishing is obvious from (\ref{adsen}), 
dropping the Einstein contribution : E is then proportional to 
$(D\a+\b)$. In addition to its zero-energy flat vacuum, the
(A)dS branch is infinitely degenerate, having a
1-dimensional moduli space denoted by the Schwarzschild parameter $r_0$.
For example, creating larger and larger black holes costs
nothing in this theory. Of course, once an Einstein term is added,
the energy is no longer  zero.

\section{Conclusions}

We have defined the energy of generic Einstein
plus cosmological term plus quadratic gravity theories as well as pure
quadratic models in all $D$, for both asymptotically flat and
(A)dS spaces. For flat backgrounds, the
higher derivative terms do not change the form of the energy
expressions. On the other hand, for asymptotically (A)dS
backgrounds (which are usually solutions to these equations, even
in the absence of an explicit cosmological constant), the energy
expressions (\ref{fullcharge}) essentially reduce to that of
the AD formula (up to higher order corrections that vanish
for space-times that asymptotically approach (A)dS at least as fast
as Schwarzschild-de-Sitter spaces ).

Among quadratic theories, we have studied the string-inspired
EGB theory in more detail. Just like the
others, this one, in the absence
of an explicit cosmological constant in the Lagrangian, has both flat
and AdS vacua, the latter with specific (negative) cosmological constants
determined by the Newton's constant and the GB coefficient, the latter
sign being fixed from the string expansion to be positive. 
The explicit spherically symmetric black hole
solutions in this theory consist of two branches {\cite{boulware}:
asymptotically Schwarzschild spaces with a positive mass parameter
or asymptotically Schwarzschild AdS spaces
with a {\it{negative}} one. The asymptotically Schwarzschild
branch has the usual positive ADM energy. Using the compensation 
of two minus signs in the solution and in 
the correct energy definition, we noted that the AdS branch has 
likewise positive energy and that the AdS vacuum was a, stable, zero 
energy, state.

Amusingly, we instead identified a 
unique, purely quadratic theory with zero energy for all 
constant (or zero )
curvature  backgrounds. That, one such model must exist, is already clear 
from the fact that each term in
\be
I = \int d^D\, x \sqrt{-g} \Big \{
\a R^2 + \b R^2_{\mu\nu} +
\gamma (R_{\mu\nu\rho\sigma}^2 -4R^2_{\mu\nu}+R^2 ) \Big \},
\label{action2}
\ee
contributes linearly to E. The condition that (A)dS be a solution , with 
arbitrary cosmological constant $\Lambda$ is 
\be
&&(D-4) \left \{{ 1\over (D-2)}(D\a +\b) +
{\gamma(D-3)\over (D-1) } \right \}=0.
\label{condition}
\ee
In all D, the zero energy models have $(D\a+\b)=0= \gamma$ .
While we have not yet understood what this result means 
physically, we can at least argue in favor 
of its plausibility .
First, note that this model is the only one that stays 
special in all $D$, unlike 
either the conformal one, good only in $D=4$ or $R^2$, 
scale invariant also only in $D=4$.
A second argument is that this is the only quadratic theory that cannot be 
reformulated as 
Einstein plus matter \cite{jakubiec},
making it hard to expect any of the others to have no energy.

In this paper, we have only looked at constant curvature vacua, but 
there may exist 
more general vacua with some additional structure. 
One example may be Weyl gravity,
for which the most general spherically symmetric solution is 
\cite{riegert,mannheim} 
\be
-g_{00} = {1\over g_{rr}} =  1- 3ab - {(2- 3ab)b\over r} + a r - 
{r^2\over l^2} ; 
\ee
$a,b, l$ are integration constants. Birkhoff's theorem
is valid and this is the unique static external solution.
One choice of  background might be  to set $b=0$. 
This space is only asymptotically (A)dS, since for it,
$R = - {6 a\over r} + { 12 \over l^2}$. 
Our earlier remarks on the loss of visibility of matter 
source contributions to E in higher derivative theories might lead one to 
expect the $ar$ term to carry this information. However this seems not to be 
the case generically, when $a=0$ is required to solve the equations. This is 
another open question.

The framework for energy definition presented here can clearly be 
applied to models with generic higher powers of curvature 
\cite{dt}. For any such theory that 
supports constant curvature vacua-and all but monomials in scalar 
curvature do so- it is just a matter 
of turning the crank to obtain the energy.

This work was supported by National Science Foundation  grant PHY99-73935.

\section{Appendix}

Here we list some useful linearization expressions about (A)dS 
for pure quadratic terms, using the conventions of Sec. 2, 
barred quantities refer to the background:
\be
&&\delta ( R_{\mu\rho\nu\sigma}R^{\rho\sigma}) =
{ 2\Lambda \over D-1}R^L_{\mu\nu} +
{2\Lambda \over (D-1)(D-2)}\bar{g}_{\mu\nu}R_L +
{4\Lambda^2 \over (D-2)^2(D-1)}h_{\mu\nu} \nonumber \\ 
&&\delta ( R_{\mu\rho\sigma\alpha}R_\nu\,^{\rho\sigma\alpha})=
{8\Lambda \over (D-1)(D-2)}R^L_{\mu\nu}
-{8\Lambda^2 \over (D-2)^2(D-1)}h_{\mu\nu} \nonumber \\
&& \delta ( R_{\mu\rho\sigma\alpha}R^{\mu\rho\sigma\alpha})=
{8\Lambda \over (D-1)(D-2)}R_L \nonumber \\
&&\delta(RR_{\mu \nu})= { 2 D\Lambda \over D-2} R^L_{\mu\nu} +
{2\Lambda \over (D-2)}\bar{g}_{\mu\nu}R_L \nonumber  \\
&& \delta( R_\mu^\sigma R_{\nu \sigma}) =
{ 4 \Lambda \over D-2} R^L_{\mu\nu} - {4 \Lambda^2 \over (D-2)^2}h_{\mu \nu} 
\nonumber \\
&&\delta(R_{\mu\nu}^2) = {4\Lambda \over D-2}R_L \nonumber \\
&&\delta( R_{\tau\lambda\rho\sigma}^2 -4R^2_{\sigma\rho}+R^2 )=
{4\Lambda (D-3)\over D-1}R_L \nonumber \\
&& R^L_{\mu\sigma\nu\rho}\bar{g}^{\sigma\rho} = R^L_{\mu\nu} -
{2\Lambda \over (D-1)(D-2)}( h_{\mu\nu} -\bar{g}_{\mu\nu}h ). \nonumber 
\ee
Finally, we compute the GB density of a cosmological space:
\be 
\bar{ R}_{\tau\lambda\rho\sigma}^2 -4{\bar{R}^2_{\sigma\rho}+\bar{R}^2} =
{4D\Lambda^2(D-3)\over (D-2)(D-1)}.\nonumber
\ee

\myend